\documentclass[preprint]{aastex}
\usepackage{graphicx}
\usepackage{amsmath}
\usepackage{amsfonts}
\usepackage{amssymb}
\usepackage{natbib}
\usepackage{textcomp}
\usepackage{gensymb}
\usepackage{url}
\shortauthors{Chitta et al.}

\begin{document}

\title{Dynamics of the solar magnetic bright points derived from their horizontal motions}

\author{LP. Chitta$^{1,3}$, A. A. van Ballegooijen$^1$, L. Rouppe van der Voort$^2$, E. E. DeLuca$^1$, and R. Kariyappa$^3$}
\affil{$^1$Harvard-Smithsonian Center for Astrophysics, 60 Garden Street MS-15, Cambridge, MA 02138, USA}
\affil{$^2$Institute of Theoretical Astrophysics, University of Oslo, P.O. Box 1029, Blindern, NO-0315 Oslo, Norway}
\affil{$^3$Indian Institute of Astrophysics, Bangalore 560 034, India}
\begin{abstract}
The sub-arcsec bright points (BP) associated with the small scale magnetic fields in the lower solar atmosphere are advected by the evolution of the photospheric granules. We measure various quantities related to the horizontal motions of the BPs observed in two wavelengths, including the velocity auto-correlation function. A 1 hr time sequence of wideband H$\alpha$ observations conducted at the \textit{Swedish 1-m Solar Telescope} (\textit{SST}), and a 4 hr \textit{Hinode} \textit{G}-band time sequence observed with the Solar Optical telescope are used in this work. We follow 97 \textit{SST} and 212 \textit{Hinode} BPs with 3800 and 1950 individual velocity measurements respectively. For its high cadence of 5 s as compared to 30 s for \textit{Hinode} data, we emphasize more on the results from \textit{SST} data. The BP positional uncertainty achieved by \textit{SST} is as low as 3 km. The position errors contribute 0.75 km$^2$ s$^{-2}$ to the variance of the observed velocities. The \textit{raw} and \textit{corrected} velocity measurements in both directions, i.e., $(v_x,v_y)$, have Gaussian distributions with standard deviations of $(1.32,1.22)$ and $(1.00, 0.86)$ km s$^{-1}$ respectively. The BP motions have correlation times of about $22 - 30$ s. We construct the power spectrum of the horizontal motions as a function of frequency, a quantity that is useful and relevant to the studies of generation of Alfv\'en waves. Photospheric turbulent diffusion at time scales less than 200 s is found to satisfy a power law with an index of 1.59. 
\end{abstract}

\maketitle
\clearpage
\section{Introduction}
The discrete and small scale component of the solar magnetic field is revealed in the high spatial resolution observations of the Sun. Ground based observations~\citep{1983SoPh...85..113M, 1985SoPh..100..237M, 1995ApJ...454..531B} show clusters or a network of many bright points (hereafter BPs) in the inter-granular lanes, with each individual BP having a typical size of $100-150$ km. These BPs are known to be kilogauss flux tubes in the small scale magnetic field (SMF), and are extensively used as proxies for such flux tubes~\citep[][see ~\citet{2009SSRv..144..275D} for a review on the SMF]{1968SoPh....5..442C, 1973SoPh...32...41S, 1985SoPh...95...99S, 1987ApJ...317..892T}. High cadence observations and studies show that magnetic BPs are highly dynamic and intermittent in nature, randomly moving in the dark inter-granular lanes\footnote{See an accompanying movie of \textit{SST} data used in this study. Also see TiO movies at \url{http://www.bbso.njit.edu/nst_gallery.html}, and \textit{G}-band movies at \url{http://solar-b.nao.ac.jp/QLmovies/index_e.shtml.}}. These motions are mainly due to the buffeting of granules. The SMF is passively advected to the boundaries of supergranules creating the magnetic network in the photosphere. 

Earlier works by several authors have reported mean rms velocities of magnetic elements in the order of a few km s$^{-1}$.  With the ground based observations of the granules at 5750~\AA~\ (white light),~\citet{1994A&A...283..232M} have identified many network BPs with turbulent proper motion and a mean speed of 1.4 km s$^{-1}$.~\citet{1996ApJ...463..365B} have used \textit{G}-band observations of the photosphere and  found that the \textit{G}-band BPs move in the intergranular lanes at speeds from 0.5 to 5 km s$^{-1}$.~\citet{1998ApJ...495..973B} observed the flowfield properties of the photosphere by comparing the magnetic network and nonmagnetic quiet Sun. They show that the convective flow structures are smaller and much more chaotic in magnetic region, with a mean speed of 1.47 km s$^{-1}$ for the tracked magnetic BPs. With the \textit{G}-band and continuum filtergrams,~\citet{1998ApJ...509..435V} used an object tracking technique and determined the autocorrelation function describing the temporal variation of the bright point velocity, with a correlation time of about 100 s. Correcting for measurements errors,~\citet{2003ApJ...587..458N} measured a 0.89 km s$^{-1}$ rms velocity of BPs. Advances in the ground based observations like rapid high cadence sequences with improved adaptive optics to minimize seeing effects, and also the space based observations at high resolutions, continued to attract many authors to pursue BP motion studies. For example,~\cite{2010A&A...511A..39U} used space based  \textit{Hinode} \textit{G}-band images to measure BP velocities and their lifetime. The BP motions can be used to measure dynamic properties of magnetic flux tubes and their interaction with granular plasma. Photospheric turbulent diffusion is one such dynamical aspect that can be derived consequently from the BP random walk.~\citet{2011A&A...531L...9M} measured a diffusion constant of 195 km$^2$ s$^{-1}$ from the BP random walk and their dispersion.~\citet{2011ApJ...743..133A} studied photospheric diffusion at a cadence of 10 s with high resolution TiO observations of a quiet Sun area. They found a super-diffusion regime, satisfying a power law of diffusion with an index $\gamma=1.53$, which is pronounced in the time intervals $10-300$ s.

The implications of these magnetic random walk motions have been found very fruitful recently. Such motions are capable of launching magneto-hydrodynamic (MHD) waves~\citep{1981A&A....98..155S}, which are potential candidates for explaining the high temperatures observed in the solar chromosphere and corona. For example, a three-dimensional MHD model developed by~\citet{2011ApJ...736....3V} suggests that random motions inside BPs can create Alfv\'{e}n wave turbulence, which dissipates the waves in a coronal loop~\citep[also see][]{2012ApJ...746...81A}. Observations by~\citet{2007Sci...318.1574D, 2009Sci...323.1582J, 2011Natur.475..477M} provide strong evidence that the Alfv\'enic waves (which are probably generated by the BP motions), have sufficient energy to heat the quiet solar corona. To test theories of chromospheric and coronal heating, more precise measurements of the velocities and power spectra of BP motions are needed.

\citet{2003ApJ...587..458N} worked on the precise measurements of BP positions, taking into account the measurement errors. The auto-correlations derived by them for the $x-$ and $y-$ components of the BP velocity using high spatial resolution and moderate cadence of 30 s observations gave a correlation time of about 60 s, which is twice the cadence of the observations. This suggests an over-estimation of correlation time and an under-estimation of the rms velocity power, with a significant hidden power in the time scales less than 30 s, and thus warranting for observations at even higher cadence. This is important because, the measured power profile, which is the Fourier transform of the auto-correlation function, gives us an estimation of the velocity amplitudes and energy flux carried by the waves that are generated by the BP motions in various and especially at higher frequencies.

In this study, we use 5 s cadence wideband H$\alpha$ observations from the Swedish 1-m Solar Telescope to track the BPs and measure their rms velocities. For comparison, we also use a 30 s cadence \textit{G}-band observational sequence from Solar Optical Telescope onboard \textit{Hinode}. These independent and complementary results take us closer to what could be the true rms velocity and power profile of the lateral motions of the BPs. The details of the datasets used, analysis procedure, results and their implications are discussed in the following sections.
\section{Datasets \label{data}}
In this study, we have analyzed time sequence of intensity filtergrams with 5 and 30 s cadence. A brief description of the observations is given below.
  
\begin{figure}
\begin{center}
\includegraphics[width=0.7\textwidth]{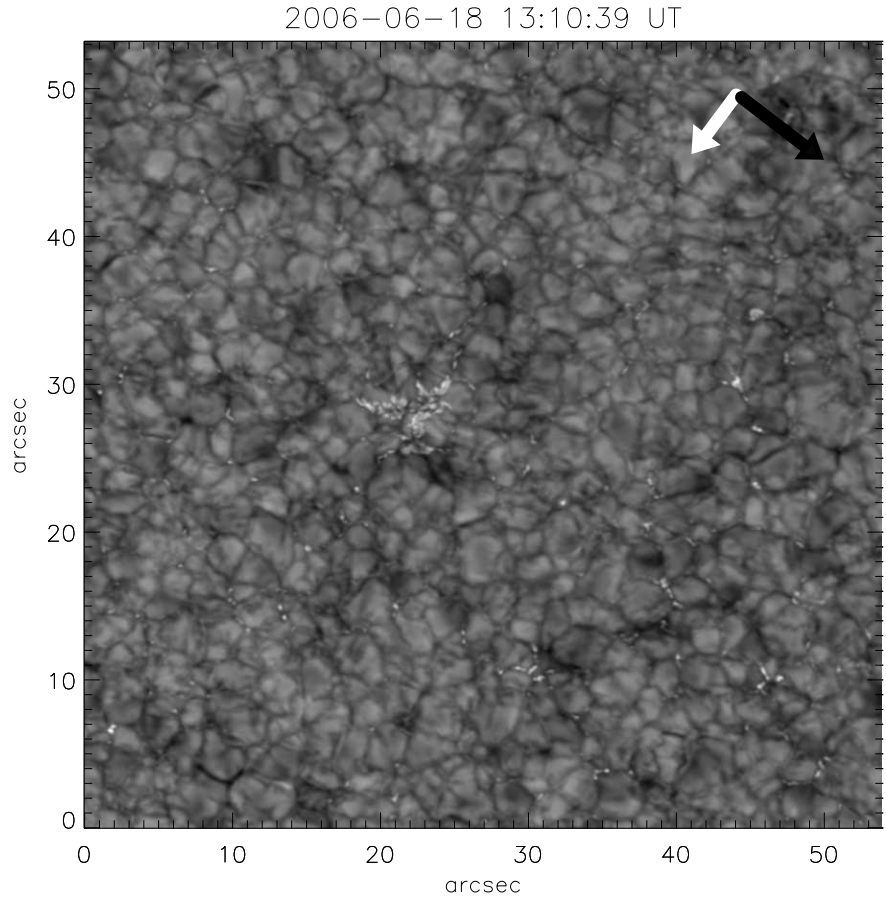}
\caption{First image from the time sequence of \textit{SST} wideband H$\alpha$ observations at 13:10 UT on 18 June 2006. Black arrow is pointing towards solar north and the white arrow is towards disk center. \label{sst-fov}}
\end{center}
\end{figure}

\textit{5 sec data}: These observations were obtained on 18 June 2006, with the \textit{Swedish 1-m Solar Telescope} (\textit{SST}; Scharmer et al. 2003a) on La Palma, using the adaptive optics system~\citep[AO;][]{2003SPIE.4853..370S} in combination with the Multi-Object Multi-Frame Blind Deconvolution~\citep[MOMFBD;][]{2005SoPh..228..191V}  image restoration method under excellent seeing conditions. The target area is a quiet Sun region away from disk center at $(x,y)=(-307'', -54'')$ and $\mu=0.94$ (see Figure~\ref{sst-fov}). The time sequence is of one hour duration starting at 13:10 UT. Here we analyze images from the wideband channel of the Solar Optical Universal Polarimeter~\citep[SOUP;][]{1981OptEn..20..815T} which received 10\% of the light before the SOUP tunable filter but after the SOUP prefilter~\citep[see][for the optical setup of the instrument]{2007ApJ...655..624D}. The prefilter was an FWHM$=$8 \AA~wide interference filter centered on the H$\alpha$ line. The SOUP filter was tuned to the blue wing of H$\alpha$ at -450 m\AA~but that data is not considered here. On the wideband channel there were 2 cameras (running at 37 frames per sec) positioned as phase-diversity pair -- one in focus and one camera 13.5 mm out of focus. the data from the two cameras has been processed with the MOMFBD restoration method in sets of 5 seconds, creating a 5 s cadence time sequence with a total of 720 images. After MOMFBD processing, the restored images were de-rotated to account for the field rotation due to the altazimuth mount of the telescope. Furthermore, the images were aligned using cross-correlation on a large area of the field of view (FOV) . The images were then clipped to $833 \times 821$ pixels (with ~$0.065''$ per pixel), to keep the common FOV (the CCDs have $1024 \times1024$ pixels, some pixels are lost after alignment between focus and defocus cameras).

For a reference direction, the solar north in the \textit{SST} time sequence is found by aligning an earlier \textit{SST} observation of that day of a magnetogram of an active region (AR) to a full disk \textit{SOHO}/MDI magnetogram (the AR was just outside the MDI High Res region). From that comparison, we fix the direction of solar north and disk center (black and white arrows respectively in Figure~\ref{sst-fov}). Though we do not rotate the images to match the solar north during our analysis, the angles are taken into account at a later stage to correct for the projection effects in the velocity measurements. 

\textit{30 sec data}: We use \textit{G}-band filtergrams observed with Solar Optical Telescope (SOT) on board \textit{Hinode} \citep{2007SoPh..243....3K, 2008SoPh..249..167T}, on 14 April 2007. The observations were made for a duration of 4 hr, with a 30 s cadence in a FOV of $55'' \times 55''$ (~$0.05''$ per pixel; 1024 pixels in both $x$ and $y$ directions), near disk center. The images were processed using standard procedures available in the \textit{solarsoft} library. 
\section{Procedure}
In this section, we briefly describe the method of determining the BP positions, and the velocity measurements through the correlation tracking.

\subsection{BP Positions \label{positions}}
We manually select the BPs to estimate their position to a sub-pixel accuracy. We consider the coordinates of maximum intensity of a given BP to be the position of that BP and the method for measuring these positions involves two steps. In the first step, we visually identify a BP and it is selected for analysis for a period during which it is clearly distinguishable from the surrounding granules.  On an average, we follow a BP for about $3 - 5$ min. The BPs with elongated shapes are not considered for the analysis. Also, we stop following a BP if it is substantially distorted or elongated from its initial shape. Though time consuming, manual selection gives a handle on the validity of the positional accuracy of a BP from frame to frame. At each time step, using a cursor, an  approximate location ($x^\prime_{app}, y^\prime_{app}$) of a particular BP is fed to an automated procedure to get its accurate position, which is step two in our method. 

Step two is completely an automated procedure. Here, we use a surface interpolation technique to get a precise position of that BP (to a sub-pixel accuracy).  Approximate position from the previous step is used to construct a grid of $5 \times 5$ pixels covering the full BP (with ($x^\prime_{app}, y^\prime_{app}$) as the center of that grid). Now, our procedure fits a 2-D, $4^{th}$ degree surface polynomial to that grid (using $SFIT$, an IDL procedure); interpolates the fit to one-hundredth of a pixel; returns the fine location of its peak ($\delta x^\prime, \delta y^\prime$) within that grid and finally stores the accurate position ($x^\prime_{BP}, y^\prime_{BP}$) of that BP (which is the sum of its approximate and fine positions ($x^\prime_{app}+\delta x^\prime, y^\prime_{app}+\delta y^\prime$)), for further analysis. Therefore, the position of a BP with index $j$ in a frame $i$ is given by
\begin{equation}
(x^\prime_{BP},y^\prime_{BP})^{j}_{i} = (x^\prime_{app}+\delta x^\prime, y^\prime_{app}+\delta y^\prime)^{j}_{i},
\end{equation}
and all the coordinates till this point are relative to the lower left corner of the image.

\subsection{Reference Frame \label{corr}}
Though the positional measurements of BPs as described in Section~\ref{positions} are accurate, they cannot be directly used to measure the velocities as there are artificial velocity sources viz. the instrumental drifts, seeing variations, jittery motions and also the solar rotation, which vectorially add to BP velocities and thus are required to be removed from the analysis. While  \textit{Hinode} (space based) data is not subjected to seeing variations, \textit{SST} (ground based) data has been corrected for seeing as described in Section~\ref{data}. Further, we need to correct for instrumental drifts, jitters and solar rotation. Calculating the offsets between the  successive images is necessary to remove these artificial velocities. In this section, we describe the method of our cross correlation analysis used to co-align the images.

Cross correlation ($C$) of two images $f(x,y)$ and $g(x,y)$ is defined as 
\begin{equation}
C = \frac{1}{k-1} \sum_{x,y}\frac{(f(x,y) - \overline{f})(g(x,y) - \overline{g})}{\sigma_f \sigma_g},
\end{equation}
where, $\overline{f}$ [$\overline{g}$] and $\sigma_f$ [$\sigma_g$] are mean value and standard deviation of $f(x,y) $ [$g(x,y)$] respectively, $k$ is the number of pixels in each image, for normalization.  With the above definition of cross correlation, to get the offsets between two images, we need to shift one image with respect to the other (in both $x$ and $y$ directions ) and find at what offsets (independent in $x$ and $y$) the correlation function attains the maximum value. In general, for shifts of $-l$ to $+l$, the cross correlation is a 2-D function with $2l+1$ rows and columns. Let $l_x$ and $l_y$ be the coarse offsets between the two images in $x$ and $y$ directions, respectively, such that the cross correlation reaches its maximum value: $max(C) = C(l_{x},l_{y})$, where $-l < l_{x}, l_{y} < l$. To get the sub-pixel offsets, the fine offsets ($\delta l_{x}, \delta l_{y}$) are calculated. The method is similar to finding the fine position of BP by using a $5 \times 5$ pixel grid but now about ($l_x,l_y$) of $C$.

Instead of cross correlating every successive image with its previous one, we keep a reference image for about 200 s, i.e., a frame $i$ taken at time $t$ ($i_{t}$) is used as a reference for the subsequent frames till $t + 200~\text{s}$ ($i_{t + 200}$) for cross correlation. Therefore the \textit{5 sec} (\textit{SST}) and the \textit{30 sec} ( \textit{Hinode}) data have about 40 and 7 images respectively in each set. By keeping the last image of a set equal to the first image in its next set, we can co-align different sets. In this way, the accumulation of errors in the offsets can be minimized.

\begin{figure}
\begin{center}
\includegraphics[width=\textwidth]{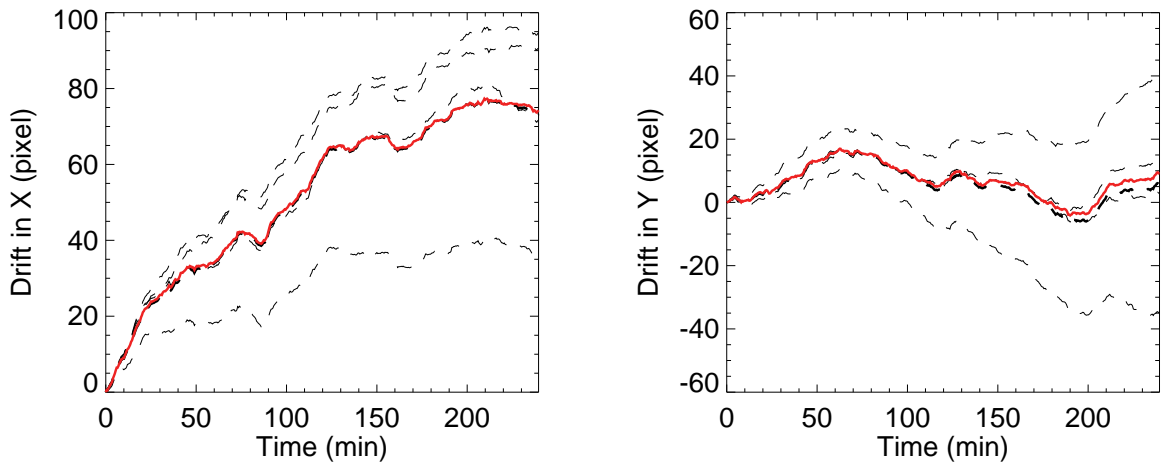}
\caption{Illustration of different offsets seen in the full and partial FOV of \textit{Hinode} data. Dashed curves in the left (right) panel show the drifts in the $x-$ ($y-$) direction of the four selected quadrants. Thick dashed profile is the average of four dashed curves. Thick red profile is the drift of the full FOV (see text for details). \label{drifts}}
\end{center}
\end{figure}

With the above background on co-aligning images to find various drifts, we present the results of drifts found in  \textit{Hinode} data.  As an illustration, we divide the full (i.e., $55'' \times 55'' \times 4$~hr)  \textit{Hinode} time sequence into 4 quadrants with $27.5'' \times 27.5'' \times 4$~hr each. Further, we do correlation tracking (as described above by keeping 7 frames per set) on each quadrant separately and plot the results in Figure~\ref{drifts}. The four dashed lines in the left and the right panels are the offsets in $x-$ and $y-$ directions respectively, the thick dashed line in each panel is the average of the offsets (i.e., average of four dashed lines), and the solid red curve is the offset obtained by considering the full FOV. Clearly, in each quadrant, the offsets have a trend similar to that of the full FOV (solid red curve) and an additional component of their own. This additional component is probably the real velocity on the Sun due to flows with varying length scales (for example, super-granular, meso-granular and granular) and with flow directions changing over areas of a few tens of arcsec$^2$ on the Sun.
 
In this paper we are mainly interested in the dynamics of the BPs relative to their \textit{local} surroundings, as granulation flows will have a dominant effect on the BP velocities and their variations on short time scales. Hence, we consider a $5'' \times 5''$ area about the BP as a reference frame for that BP (i.e., keeping the BP in the center of the $local$ area). The cross correlation is performed on this $5'' \times 5''$ area instead of the full FOV to get the offsets, which are subtracted from the $(x^\prime_{BP},y^\prime_{BP})^{j}_{i}$. The BP positions corrected for offsets are now given by 
\begin{equation}\label{eq:bpc}
(x^\prime_{BPC},y^\prime_{BPC})^{j}_{i} = (x^\prime_{BP},y^\prime_{BP})^{j}_{i} - (l_{x} + \delta l_{x}, l_{y} + \delta l_{y})^{j_{local}}_{i},
\end{equation}
where $j_{local}$ represents the $local$ area of BP$^j$.

In the case of the \textit{SST} data, the observations are off disk center at $(-307'',-54'')$, which corresponds to a heliocentric angle of $\arccos(0.94)$. This will introduce a projection effect on the measured horizontal velocities in both $x^\prime-$ and $y^\prime-$ directions and needs to be corrected. To do this, the coordinate system $(x^\prime,y^\prime)$ defined by the original \textit{SST} observations, is rotated by $45\degree$ in the anti-clockwise direction. Now the image plane is oriented in E-W (parallel to equator, new $x-$) and N-S (new $y-$)  directions. Further, the E-W coordinate is multiplied by a factor of $0.94^{-1}$.  Hence the new coordinate system $(x,y)$ is given by
\begin{eqnarray}\label{eq:bpcn}
x &=& (x^\prime\cos 45\degree +  y^\prime\sin 45\degree) \times \frac{1}{0.94}, \nonumber \\
y &=& (-x^\prime\sin 45\degree +  y^\prime\cos 45\degree).
\end{eqnarray}
\textit{SST} BP positions $(x^\prime_{BPC},y^\prime_{BPC})^{j}_{i}$, as measured from Equation~\eqref{eq:bpc}, are remapped to $(x_{BPC},x_{BPC})^{j}_{i}$, using the above coordinate transformations\footnote{Note that the transformations in Equation~\eqref{eq:bpcn} are only to modify the \textit{SST} BP positions and in the rest of the paper, we use $(x,y)$ for the remapped $(x^\prime,y^\prime)$ of \textit{SST} and $(x,y)$ of  \textit{Hinode}.}.  

 \section{Results \label{results}}
\begin{figure}
\includegraphics[width=\textwidth]{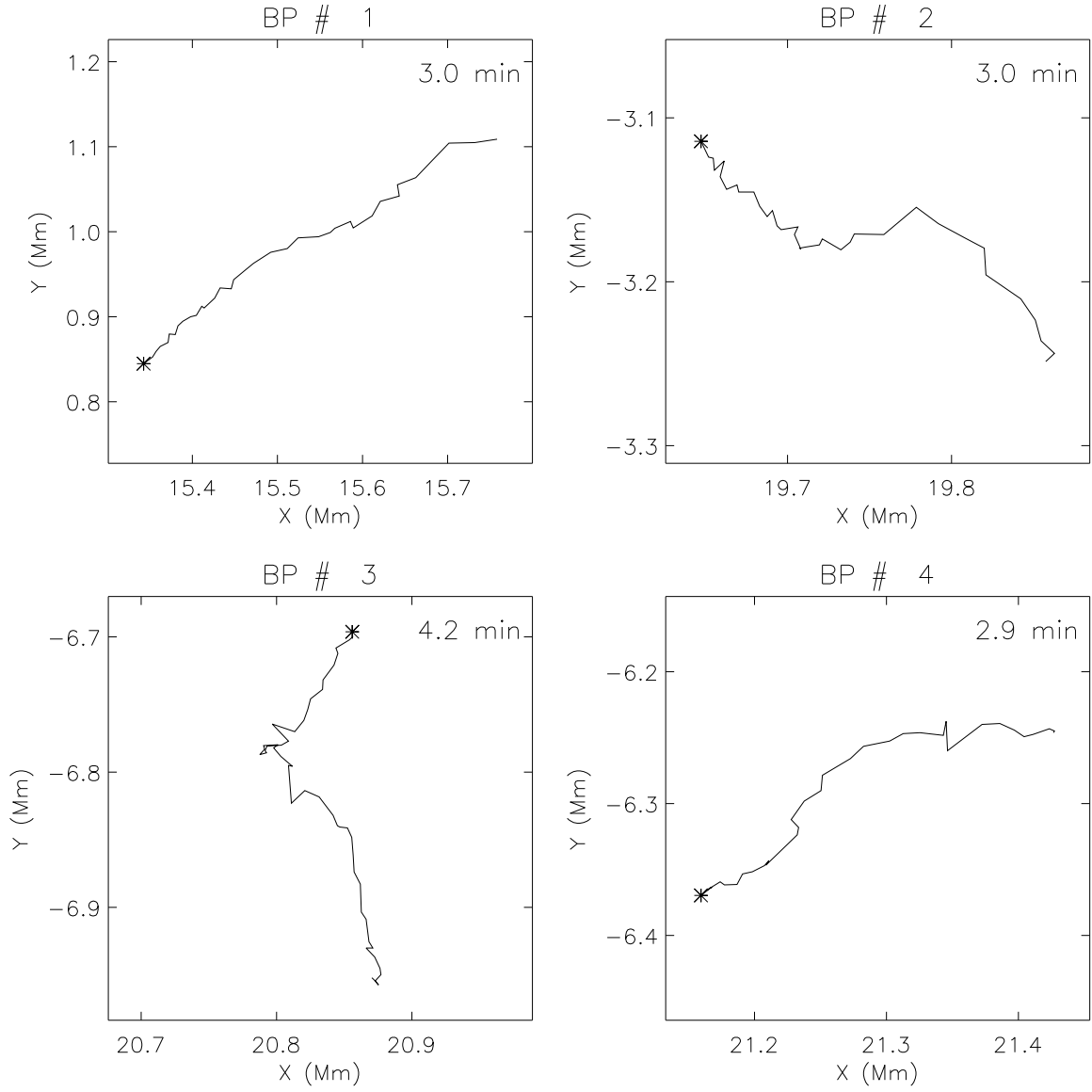}
\caption{Examples of the paths of four BPs taken from the \textit{SST} data. Initial position of each BP is marked with a star. Time shown at the top right corner in each panel is the duration for which respective BP is followed.\label{track}}
\end{figure}

\begin{figure}
\includegraphics[width=\textwidth]{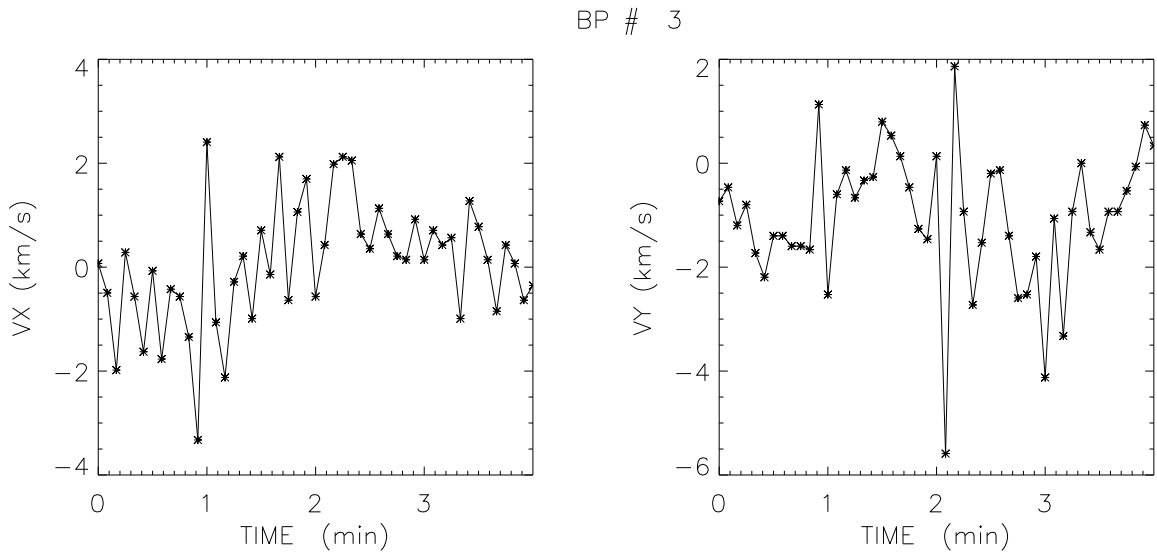}
\caption{Velocities $v_{x}$ and $v_{y}$ as a function of time for a typical BP (shown here for BP\#3, see Figure~\ref{track} for the track of BP\#3).\label{vel}}
\end{figure}

In this section, we present various results in detail giving more emphasis on the \textit{SST} results. We have selected 97 \textit{SST} BPs with $\sim$3800 individual velocity measurements\footnote{Similarly, we have identified 212 \textit{Hinode} BPs with 1950 individual velocity measurements.}. Figure~\ref{track} shows the paths of four individual \textit{SST} BPs. Some of the BPs move in a relatively smoother path while some exhibit  very random motions to the shortest time steps available. BPs drift about a few hundred km in a few minutes. The instantaneous velocity $(v_x, v_y)^j_{i+1}$ of a BP is given by $(x_{BPC}, y_{BPC})^j_{i+1} - (x_{BPC}, y_{BPC})^j_{i} $, multiplied by a factor to convert the units of measured velocity to km s$^{-1}$ (9.4 in case of \textit{SST} which is, image scale of \textit{SST} in km divided by the time cadence in sec). Figure~\ref{vel} shows the plot of such velocities as a function of time for BP\#3 (path of BP\#3 is shown in the lower left panel of Figure~\ref{track}). Usually, the changes in the velocity are gradual in time but, sometimes we do see sudden and large changes in the magnitude and direction of the velocity (for example at 1 min in $v_x$ and at 2 min in $v_y$ in Figure~\ref{vel}). Note that a large change of velocity of one sign is followed immediately by a change of the opposite sign, so the net change in position is not very large. This suggest that these changes are due to errors in the positional measurements. A position error at one time will affect the velocities in the intervals immediately before and after that time. In the following we will assume that such changes in velocity are due to measurement errors. However, we cannot rule out that some of these changes are due to real motions on the Sun on a time scales less than 5 seconds.

\begin{figure}
\includegraphics[width=\textwidth]{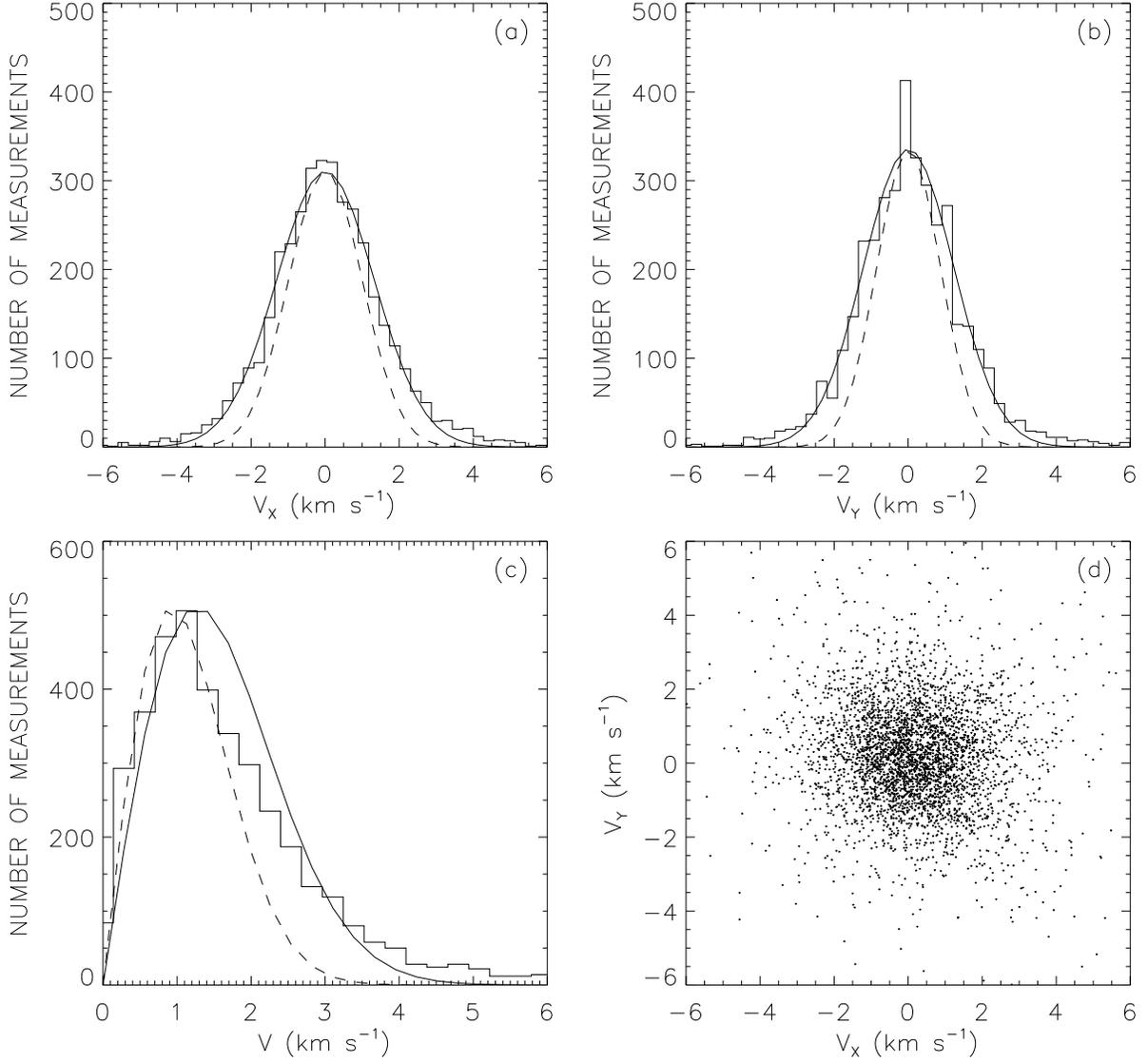}
\caption{Histograms of measured BP velocities: (a) $v_x$, (b) $v_x$, (c) $v=\sqrt{v^2_x+v^2_y}$. Solid Gaussians in the top panels are fits to the histograms. Dashed Gaussians in panels (a), (b), and dashed Rayleigh profile in panel (c), are the new distributions of velocities after correcting for the measurement errors (see text for details); (d) shows $v_x$ plotted against $v_y$. \label{hist}}
\end{figure}

The means and standard deviations of $v_x$ and $v_y$ are listed in Table~\ref{tab:tab1} (first line). Histograms of the distribution of velocities $v_x$, $v_y$ and $v = \sqrt{v_x^2 + v_y^2}$, are shown in Figure~\ref{hist} (panels (a), (b) and (c) respectively). Solid lines in panels (a) and (b) are Gaussian fits to the histograms with \textit{raw} standard deviations ($\sigma_{v,r}$) of 1.32 and 1.22 km s$^{-1}$. A scatter plot of $v_x$ against $v_y$ is shown in panel (d), which is symmetric in the $v-$space. However, a small non-zero and positive mean velocity of about 0.2 km s$^{-1}$ is noticed, suggesting that there is a net BP velocity with respect to the 5 arcsec boxes that we used as reference frames. Values of the mean and rms velocities as determined from the fits are also listed in Table~\ref{tab:tab1} (second line). These distributions are a mix of both true velocities and measurement errors. 

\begin{figure}
\includegraphics[width=\textwidth]{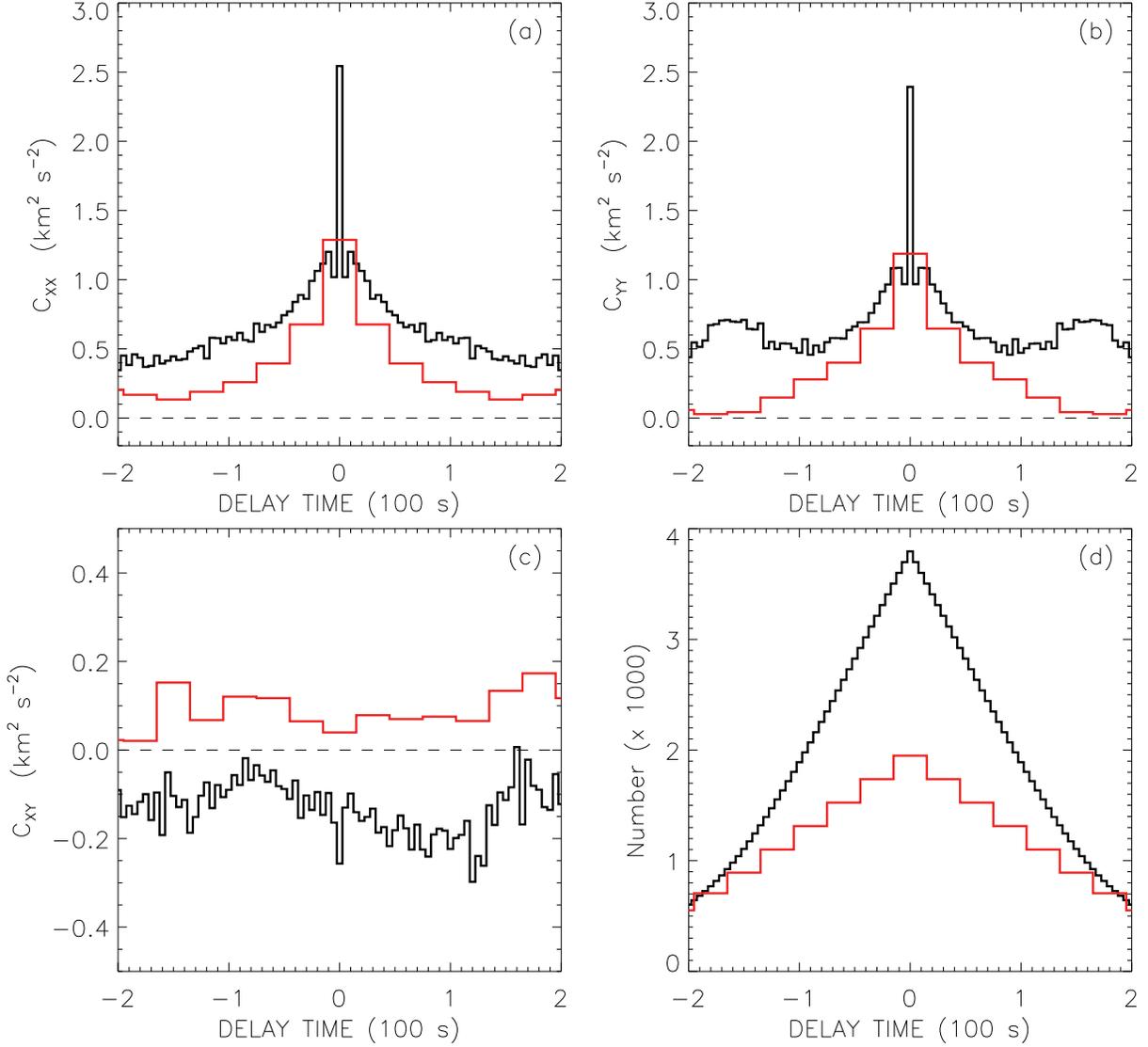}
\caption{Correlation functions of BP velocity $v_x$ and $v_y$. (a) Observed auto-correlation $c_{xx,n}$ as a function of delay time $t$ (black: \textit{SST}, red: \textit{Hinode}). (b) Similar for the observed auto-correlation $c_{yy,n}$. (c) Cross-correlation $c_{xy,n}$ as function of delay time. (d) Number of measurements per bin used in panels (a), (b), and (c). \label{auto}}
\end{figure}

We can gain more insight into the the dynamical aspects of the BP motions by studying their observed velocity correlation function $c(t)$, defined as
\begin{eqnarray}
c_{xx,n} &=& \langle v_{x,i}^j v_{x,i+n}^j \rangle,~\ c_{yy,n} = \langle v_{y,i}^j v_{y,i+n}^j \rangle \label{eq:auto}\\
c_{xy,n} &=& \langle v_{x,i}^j v_{y,i+n}^j \rangle, \label{eq:cross}
\end{eqnarray}
where $c_{xx,n}$, $c_{yy,n}$ are the auto-correlations, and $c_{xy,n}$ is the cross-correlation of $v_x$ and $v_y$, $n$ is the index of the delay time. $\langle\text{\dots}\rangle$ denotes the average over all values of the time index $i$ and BP index $j$ but for a fixed value of $n$. These results are shown in Figure~\ref{auto}. Top left and right panels are the plots of $c_{xx}$ and $c_{yy}$ respectively. Black curves are for the \textit{SST} whereas the red curves show the \textit{Hinode} results for comparison. Both the \textit{SST} and \textit{Hinode} results are consistent for delay times $< 1$ min. However, the \textit{Hinode} auto-correlations quickly fall to lower values. This is mainly a statistical error, since we do not have a large number of measurements in the case of \textit{Hinode}. Focusing on periods $< 1$ min, it is clear from the auto-correlation plots that the core of the \textit{Hinode} data within $\pm30$ s delay time, which is sampled with three data points is now well resolved with the aid of the \textit{SST} data. Also, at shorter times, $c$ takes a cusp-like profile. Extrapolating this to delay times of the order of 1 s, we expect to see a steep increase in the rms velocities\footnote{The correlation at zero-time lag is the variance of the velocity distribution.} of the BP motions. The bottom left panel shows the cross-correlation as function of delay time. The \textit{SST} data show a small but a consistent and overall negative $c_{xy}$ while the \textit{Hinode} data show a small positive correlation. We suggest that the real cross-correlation $c_{xy} \approx 0$, and that the measured values are due to a small number of measurements with high velocities (largely exceeding the rms values). The lower right panel of Figure~\ref{auto} shows the number of measurements $N_n$ used in the correlation analysis for both the \textit{SST} and \textit{Hinode} data. To obtain good statistics we collected enough BP measurements to ensure that $N_n \gtrsim 500$ for all bins.

In the rest of the section, we describe the method of estimating the errors in the velocity measurements due to positional uncertainties by analyzing $c(t)$. Following~\citet{2003ApJ...587..458N}, we assume that the errors in the positions are uncorrelated from frame to frame and randomly distributed with a standard deviation of $\sigma_p$. Since the velocities are computed by taking simple differences between position measurements (see above), the measurement errors increase the observed velocity correlation at $n=0$ by $\Delta$ (error), and reduce the correlations at $n=\pm1$ by $-\Delta/2$  where $\Delta = 2(\sigma_p/\delta t)^2$ and $\delta t$ is the cadence (see Equation 3 in their paper). We define
\begin{equation}
\Delta_n = \begin{cases}
        \Delta  & \text{when $n=0$ }\\
        -\frac{1}{2}\Delta & \text{when $n=\pm1$}\\
        0 & \text{otherwise,}
        \end{cases} \label{eq:delta}
\end{equation}
which is valid only with our two-point formula for the velocity. Once $\Delta$ is determined, the rms values ($\sigma_{v,c}$) of the true solar velocities can be measured as $\sigma_{v,c}^2 = \sigma_{v,r}^2 - \Delta$. 

A previous study using data from the \textit{Swedish Vacuum Solar Telescope}~\citep{1998ApJ...509..435V} assumed $c(t)$ to be a Lorentzian. Here, we clearly see that $c(t)$ differs from a Lorentzian, and it can be fitted with a function $\mathfrak{C}$, which is a sum of the true correlation of solar origin ($\mathfrak{C}^\prime$) and $\Delta$, given by
 \begin{equation}
  \mathfrak{C}_n(\Delta, \tau, \kappa) =   \mathfrak{C}^\prime_n(\tau, \kappa) + \Delta_n, \label{eq:c}
    \end{equation}
    where
    \begin{equation}
 \mathfrak{C}^\prime_n(\tau, \kappa) = a + \frac{b}{1+\left (\frac{|t_n|}{\tau}\right )^\kappa} \label{eq:c1}
 \end{equation}
is a generalized Lorentzian. $\Delta$, $\tau$ (correlation time), and $\kappa$ (exponent) are the free parameters of the fit; $a$ and $b$ are the functions of $(\Delta, \tau, \kappa)$, which are determined analytically by least square minimization (see Appendix~\ref{app:appen}). We also bring to the notice of the reader that our formula for $\mathfrak{C}$ is a monotonically decreasing function of $t_n$. However, there is an unexplained increase in the observed $c_{yy}$ beyond $\pm 100$ s (panel (b) in Figure~\ref{auto}). To eliminate any spurious results due to this anomaly, we use a maximum delay time of $\pm 105$ s to fit $c$ with $\mathfrak{C}$ by minimizing the sum of the squares of their difference, as defined in Equation~\eqref{eq:xy}.

\begin{figure}
\includegraphics[width=\textwidth]{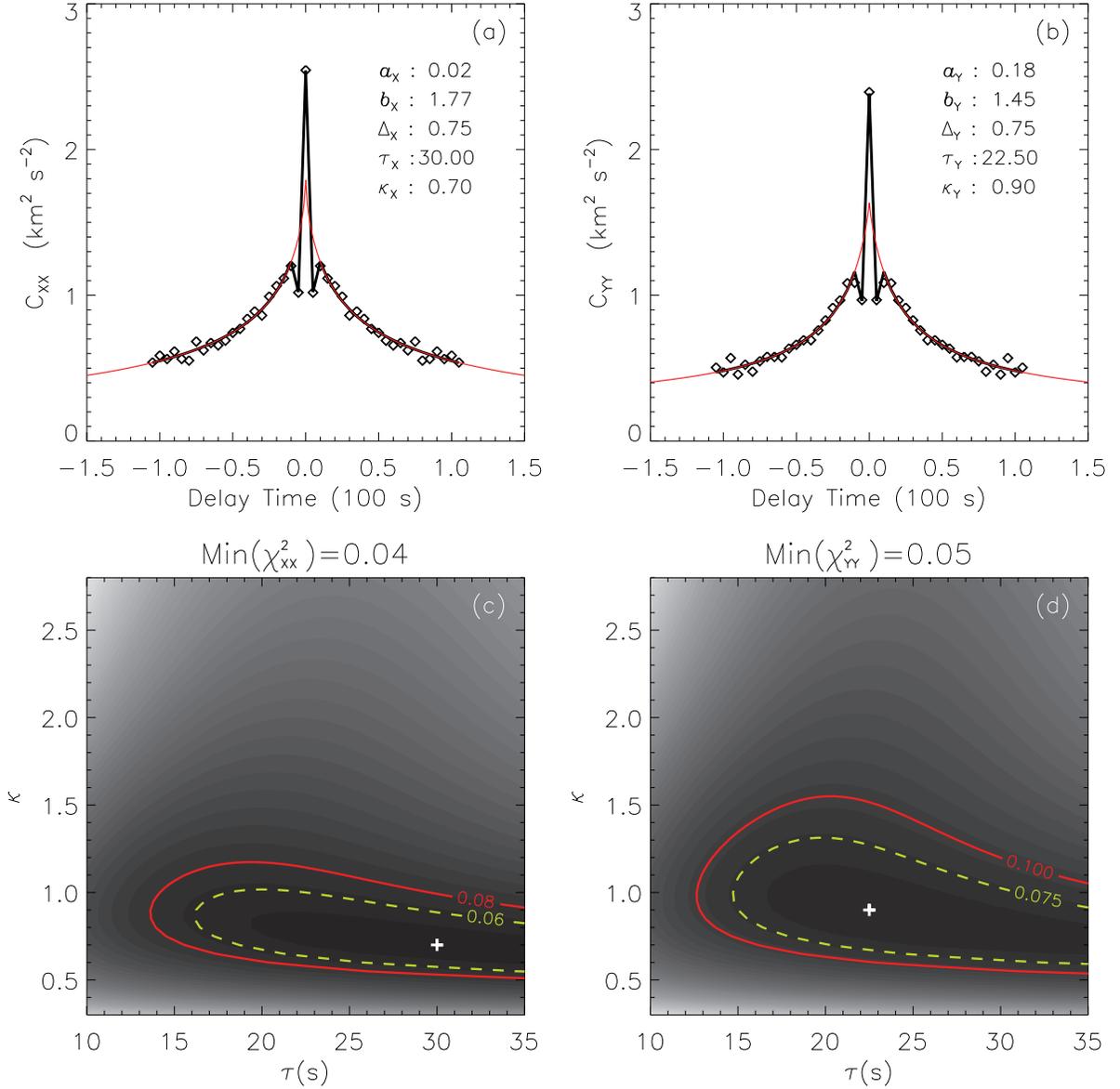}
\caption{Top: $\mathfrak{C}$ (black curve) plotted as a function of delay time with the best fit values of $a$, $b$, $\Delta$, $\tau$, and $\kappa$ obtained by minimizing the $\chi^2$ (see Appendix~\ref{app:appen}), of the observed $c$ (shown as symbols, $c_{xx}$: left, and $c_{yy}$: right, also shown as black curves in the top panels of Figure~\ref{auto}), and the modeled correlation function $\mathfrak{C}$, for a delay time of $\pm105$ s in steps of 5 s. Thin red curve is the profile of $\mathfrak{C}^\prime$. Bottom: Contour plots of $\chi^2$ as a function of $\kappa$ and $\tau$, for a value of $\Delta$ ($\Delta_x = \Delta_y = 0.75$ km$^2$ s$^{-2}$), where $\chi^2$ attains the global minimum. Plus symbol is the global minimum of $\chi^2$, dashed and solid lines are the contours of $1.5min(\chi^2)$ and $2min(\chi^2)$ respectively. \label{fit1}}
\end{figure}

The top panel in Figure~\ref{fit1} shows the results listing the best fit values of the free parameters $(\Delta, \tau, \kappa)$, $a$, and $b$ for a maximum $t_n$ of $\pm 105$ s. $\mathfrak{C}$ (black) and $\mathfrak{C}^\prime$ (thin red) are plotted as functions of the delay time over $c_{xx,n}$ (left, symbols), and $c_{yy,n}$ (right, symbols). The value of $\Delta$ where $\chi^2$ has its global minimum is found to be 0.75 km$^2$ s$^{-2}$, for both $c_{xx}$ and $c_{yy}$. The bottom panel shows the contours of $\chi^2$ as a function of $\tau$ and $\kappa$ at $\Delta = 0.75$ km$^2$ s$^{-2}$, and the $min(\chi^2)$ is denoted by plus symbols. Dashed and solid lines are the regions of 1.5 and 2 $times$ the $min(\chi^2)$ respectively. $\chi^2$ is a well bounded function for $\kappa <  2$, confirming a cusp-like profile. The correlation time is $22 - 30$ s, which is about $4 - 6$ \textit{times} the time cadence. 

Taking into account the variance in errors (i.e., $\Delta=0.75$ km$^2$ s$^{-2}$), we get $\sigma_p=3$ km, and the corrected rms velocities ($\sigma_{v,c}$) of $v_x$ and $v_y$ are now 1.00 and 0.86 km s$^{-1}$. These results are plotted as dashed curves in panels (a) and (b) of Figure~\ref{hist}, and the values are tabulated in the last row of Table~\ref{tab:tab1}. The corrected distribution of $v$ is shown as a dashed Rayleigh distribution in panel (c). With higher cadence observations, these results can be refined and modified, as $(\Delta, \tau, \kappa)$ depend on the shape of the core of $c$. By comparing the \textit{SST} and  \textit{Hinode} results, we expect that the observed $c$ probably increases rapidly below 5 s and thus changing the set of parameters to some extent.
\section{Summary and Discussion}
We studied the proper motions of the BPs using the wideband H$\alpha$ observations from the \textit{SST} and the \textit{G}-band data from \textit{Hinode}. BPs are manually selected and tracked using $5'' \times 5''$ areas surrounding them as reference frames. The quality of the \textit{SST} observations allowed us to measure the BP positions to a sub-pixel accuracy with an uncertainty of only 3 km, which is at least seven times better than the value reported by~\citet{2003ApJ...587..458N}, and comparable to the rms value of 2.7 km due to image jittering reported by~\citet{2011ApJ...743..133A}. They adopted this rms value of 2.7 km as a typical error of calculations of the BP position. We found that the horizontal motions of the BPs in $x$ and $y$ are Gaussian distributions with \textit{raw} (including the true signal and measurement errors) rms velocities of 1.32 and 1.22 km s$^{-1}$, symmetric in $v-$space, observed at 5 s cadence. The above estimate of the measurement uncertainty is obtained from a detailed analysis of the velocity auto-correlation functions. For this, we fitted the observed $c(t)$ with $\mathfrak{C}$, a function of the form shown in Equation~\eqref{eq:c}, and estimated an rms error of about 0.87 km s$^{-1}$ in $v_x$ and $v_y$. The removal of this error makes the $v_x$ and $v_y$ Gaussians narrower with new standard deviations 1.00 and 0.86 km s$^{-1}$ (a fractional change of 30\%). The total rms velocity ($v_x$ and $v_y$ combined) is 1.32 km s$^{-1}$. The correlation time is found to be in the range of $22 - 30$ s.

\begin{figure}
\begin{center}
\includegraphics{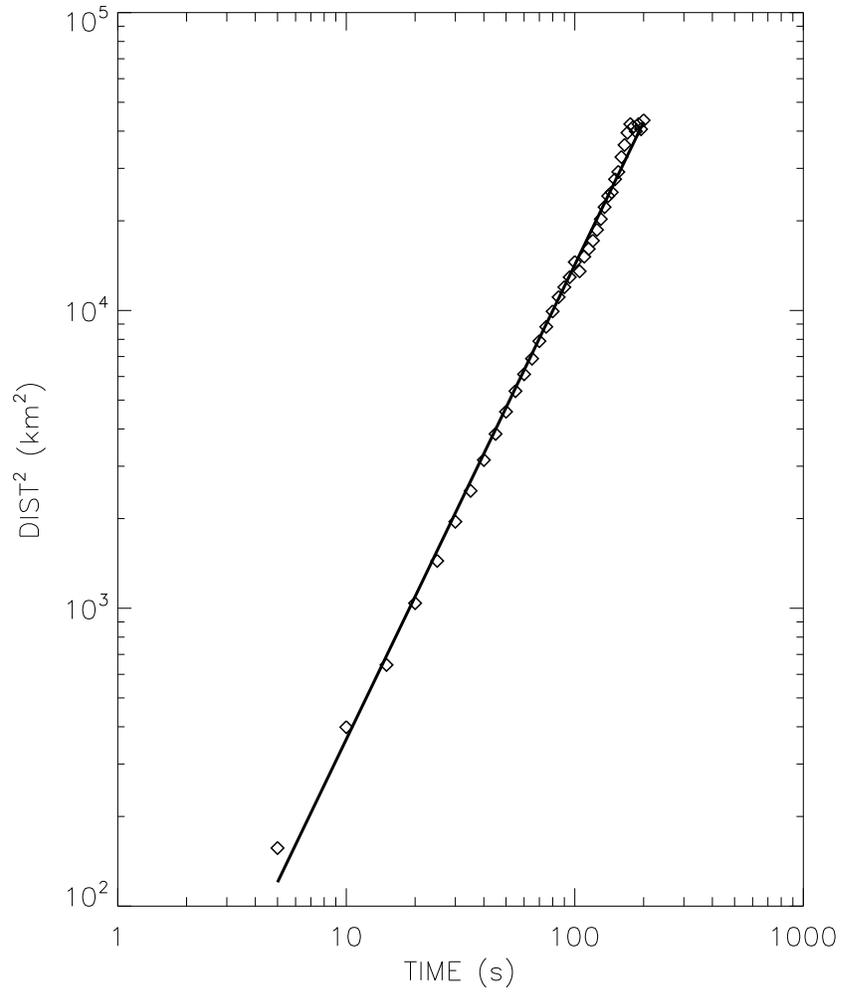}
\caption{Mean squared displacement $\langle (\Delta r)^2\rangle$ as a function of time $t$ on a log-log scale. Solid line is the least square fit of the observations (symbols), with a slope of 1.59. \label{diffusion}}
\end{center}
\end{figure}

Following is a brief note and discussion on the additional results we derive from our work. BPs are advected by the photospheric flows. Thus, taking these features as tracers, we can derive the diffusion parameters of the plasma. As BPs usually have life times of the order of minutes, the motion of these features can be used to study the nature of photospheric diffusion at short time scales. The mean squared displacement of BPs $\langle (\Delta r)^2\rangle$ as function of time is a measure of diffusion. It is suggested in the literature that $\langle (\Delta r)^2\rangle$ can be approximated as a power law with an index $\gamma$~\citep[i.e., $\langle (\Delta r)^2\rangle \sim t^\gamma$, see for example][]{1999ApJ...521..844C, 2011ApJ...743..133A}.  In Figure~\ref{diffusion}, we plot the observed $\langle (\Delta r)^2\rangle$ (symbols), for the 200 s interval on a log-log scale. Solid line is the least square fit with a slope of 1.59,  which is consistent with the value $\gamma=1.53$ found by~\citet{2011ApJ...743..133A} for quiet sun. Despite the differences in the observations (instruments and observed wavelengths), and analysis methods (identification and tracking of BPs), a close agreement in the independently estimated $\gamma$ suggests that this is a real solar signal. Both these results assert the presence of super-diffusion (i.e., $\gamma > 1$), for time intervals less than 300 s. Since most of the BPs in this study are tracked for only $3 - 4$ min, we cannot comment on the diffusion at longer times.  

Note that there is a general relationship between the mean squared displacement $\langle (\Delta r)^2\rangle$ and the velocity auto-correlation function $\mathfrak{C}^\prime$
\begin{eqnarray}
\langle (\Delta r)^2\rangle &=& \langle \left(\int\limits_0^t v_x(t')dt'\right)^2 \rangle + \langle \left(\int\limits_0^t v_y(t')dt'\right)^2 \rangle \\
                                               &=& 2\int\limits_0^t \int\limits_0^t \mathfrak{C}^\prime(t''-t')dt'dt'',
\end{eqnarray}
where we assume isotropy of the BP motions ($\mathfrak{C}^\prime_{xx}=\mathfrak{C}^\prime_{yy}=\mathfrak{C}^\prime$). For a known auto-correlation or mean squared displacement, the other quantity can be derived using the above relation.

\begin{figure}
\begin{center}
\includegraphics{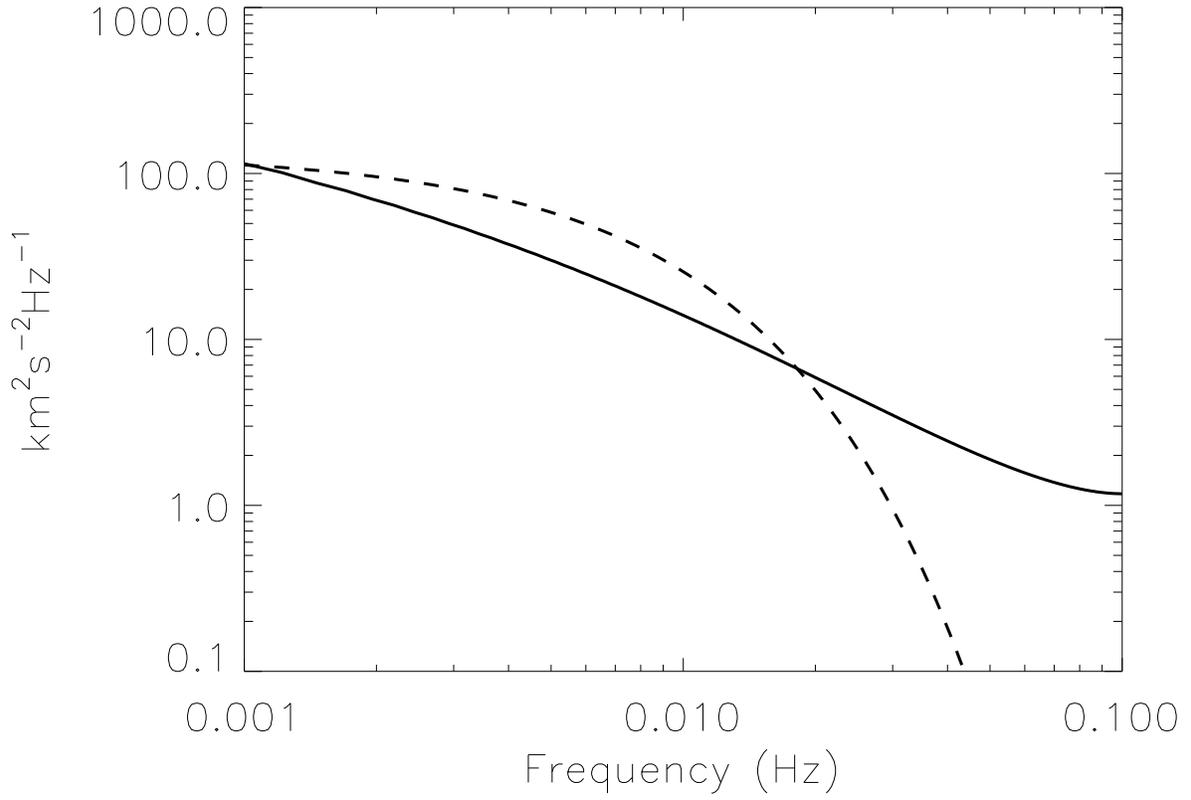}
\caption{Power spectrum of the horizontal motions of BPs as a function of frequency derived from auto-correlation function for two cases. Solid line: case (a) -- from this study. Dashed line: case (b) -- from a Lorentz profile with same $a$, $b$, and $\tau$ as in case (a) but with $\kappa=2$ (see text for details).\label{pwrsp}}
\end{center}
\end{figure}

We already saw that the horizontal motions of the BPs yield several important properties of the lower solar atmosphere. One more such important property is the possibility of the generation of Alfv\'en waves due to these motions. Here we qualitatively estimate and compare the power spectrum of horizontal motions as a function of frequency for two forms of the velocity correlation function\footnote{Fourier transform of the velocity auto-correlation is the power spectrum.}: (a) the form $\mathfrak{C}^\prime$ (Equation~\eqref{eq:c1}), obtained in this study, and (b) a Lorentzian function. For case (a) we use $a=0$, and also assume that $\mathfrak{C}^\prime_{xx,n} = \mathfrak{C}^\prime_{yy,n}$, with the parameters $b$, $\tau$, and $\kappa$ taking the mean values of $x$ and $y$. For case (b) we use a modified form of $\mathfrak{C}^\prime$ with $\kappa=2$. The other parameters ($a$, $b$, and $\tau$) are the same as in case (a).  Figure~\ref{pwrsp} shows the power spectra for the two described cases: (a) solid line, and (b) dashed line. We observe that for frequencies exceeding 0.02 Hz ($<$ 50 s), the horizontal motions generally have more power in case (a) as compared to case (b). This highlights the fact that the dynamics of the BPs at short time scales are very important. Therefore, it is highly desirable to do these observations and calculations at very high cadence. 

The measurements presented in this paper provide important constraints of models for Alfv\'en and kink wave generation in solar magnetic flux tubes. As discussed in the Introduction, such waves may play an important role in chromospheric and coronal heating. In the Alfv\'en wave turbulence model~\citep{2011ApJ...736....3V, 2012ApJ...746...81A} it was assumed that the photospheric footpoints of the magnetic field lines are moved about with rms velocity of 1.5 km s$^{-1}$, similar to the rms velocity of 1.32 km s$^{-1}$ found here. However, the models include only the internal motions of a flux tube, whereas the observations refer to the displacements of the flux tube as a whole. Clearly, to make more direct comparisons between models and observations will require imaging with high spatial resolution ($<$ 0.1 arcsec). This may be possible in the future with the Advanced Technology Solar Telescope.

In this work we presented the results of the BP motions, some of their implications and use in the context of photospheric diffusion and coronal wave heating mechanisms. We interpret the location of the intensity maximum of a BP as its position at any given time. This is certainly plausible for time periods when we begin to see the physical motion of a BP as a \textit{rigid body} due to the action of the convection on the flux tubes. But at timescales shorter than one minute, other interpretations are also plausible: the motions marked by the intensity maxima could be intensity fluctuations in an otherwise static BP. Nevertheless, these fluctuations are manifestations of some disturbances inside the BP, which are equally important and interesting to explore further. 
\clearpage
\begin{deluxetable}{cccccc} 
\tablecolumns{6} 
\tablewidth{0pc} 
\tablecaption{Properties of the velocity distributions in Figure~\ref{hist} \label{tab:tab1}} 
\tablehead{ 
\colhead{}    &  \multicolumn{2}{c}{$v_x$ (km s$^{-1}$)} &   \colhead{}   & 
\multicolumn{2}{c}{$v_y$ (km s$^{-1}$)} \\ 
\cline{2-3} \cline{5-6} \\ 
\colhead{} & \colhead{$\langle v_x \rangle$}   & \colhead{$\sigma(v_x)$}    &  
\colhead{}    & \colhead{$\langle v_y \rangle$}   & \colhead{$\sigma(v_y)$}}    
\startdata 
Histogram & 0.18 & 1.58 & & 0.19 & 1.54 \\
Gaussian Fit ($\sigma_{v,r}$)& 0.01 & 1.32 & & 0.01 & 1.22 \\
Corrected Distribution ($\sigma_{v,c}$) & 0.01 & 1.00 & & 0.01 & 0.86 \\
\enddata
\end{deluxetable} 
\clearpage
\acknowledgments  Authors thank the referee for many comments and suggestions that helped in improving the presentation of the manuscript. LPC. is a $2011-2012$ SAO Pre-Doctoral Fellow at the Harvard-Smithsonian Center for Astrophysics. The \textit{Swedish 1-m Solar Telescope} is operated on the island of La Palma by the Institute for Solar Physics of the Royal Swedish Academy of Sciences in the Spanish Observatorio del Roque de los Muchachos of the Instituto de Astrof\'{i}sica de Canarias. Funding for LPC and EED is provided by NASA contract NNM07AB07C. \textit{Hinode} is a Japanese mission developed and launched by ISAS/JAXA, collaborating with  NAOJ as a domestic partner, NASA and STFC (UK) as international partners. Scientific operation of the \textit{Hinode} mission is conducted by the \textit{Hinode} science team organized at ISAS/JAXA. This team mainly consists of scientists from institutes in the partner countries. Support for the post-launch operation is provided by JAXA and NAOJ (Japan), STFC (U.K.), NASA (U.S.A.), ESA, and NSC (Norway). This research has made use of NASA's Astrophysics Data System.

\clearpage
\appendix
\section{Determination of $a$ and $b$ \label{app:appen}}
In this section, we briefly describe a method of determining $a$ and $b$ for a set of parameters $(\Delta, \tau, \kappa)$. We define $\chi^2$ of autocorrelation functions of $v_x$ and $v_y$ as 
\begin{eqnarray}\label{eq:xy}
\chi^2_{xx}(\Delta, \tau, \kappa) &=& \sum_{n= -N}^{N} \left [c_{xx, n} - \mathfrak{C}_n(\Delta, \tau, \kappa)\right ]^2, \mbox{and} \nonumber\\
\chi^2_{yy}(\Delta, \tau, \kappa) &=& \sum_{n= -N}^{N} \left [c_{yy, n} - \mathfrak{C}_n(\Delta, \tau, \kappa)\right]^2
\end{eqnarray}
where, $c_{xx, n}$ and $c_{yy, n}$ are the observed autocorrelation values of velocities $v_x$ and $v_y$ and $\mathfrak{C}$ is a model of the correlation function given by Equation~\eqref{eq:c}.  By minimizing the $\chi^2$ with respect to $a$ and $b$ (i.e., $\frac{\partial \chi^2}{\partial a} = 0$ and $\frac{\partial \chi^2}{\partial b} = 0$, separately for $x$ and $y$), and solving the resulting system of linear equations, we have
\begin{eqnarray}\label{eq:sol}
a &=& \frac{1}{\alpha\beta^\prime-\alpha^\prime\beta}\left(\beta^\prime A - \beta B \right) \\
b &=& \frac{1}{\alpha\beta^\prime-\alpha^\prime\beta}\left(\alpha B - \alpha^\prime A \right)
\end{eqnarray}
where, 
\begin{eqnarray}
\alpha&=& 2n + 1\nonumber \\
\beta&=&\displaystyle\sum_{n= -N}^{N}\frac{1}{1+\left(\frac{|t_n|}{\tau}\right)^\kappa} \nonumber \\
\alpha^\prime&=&\beta \nonumber \\
\beta^\prime&=&\displaystyle\sum_{n= -N}^{N}\frac{1}{\left[1+\left(\frac{|t_n|}{\tau}\right)^\kappa\right]^2} \nonumber \\
A&=&\displaystyle\sum_{n= -N}^{N}c_n \nonumber \\
B&=&\displaystyle\sum_{n= -N}^{N}\left( \frac{c_n-\Delta_n}{1+\left(\frac{|t_n|}{\tau}\right)^\kappa}\right) \nonumber
\end{eqnarray}
\clearpage
\end{document}